\documentclass{momento}
\usepackage[pdftex]{graphicx}
\usepackage[spanish,activeacute]{babel}
\usepackage[latin1]{inputenc}
\usepackage{verbatim}
\usepackage[centerlast,sc,footnotesize]{caption2}
\usepackage{subfigure}
\usepackage{amsmath}

\title{ESTUDIO TEÓRICO DEL ESPECTRO DE EMISIÓN PARA UN SISTEMA MICROCAVIDAD-PUNTO CUÁNTICO EN UNA APROXIMACIÓN DE CAMPO MEDIO\\[0.5cm]
THEORETICAL STUDY OF EMISSION SPECTRUM FOR A QUANTUM DOT-MICROCAVITY SYSTEM IN A MEAN FIELD APPROXIMATION}

\author{Juan S. Rojas-Arias\suprm1, B. Rodríguez\suprm2, Herbert Vinck-Posada\suprm3}

\newcommand{\afiliacion}{\suprm {1,3} Departamento de Física, Universidad Nacional de Colombia\\ \suprm 2 Instituto de Física, Universidad de Antioquia}

\newcommand{\autor}{Juan S. Rojas-Arias, B. Rodríguez, Herbert Vinck-Posada}
\newcommand{\tema}{Estudio teórico del espectro de emisión para un sistema microcavidad-punto cuántico...}

\authorinfo{jusrojasar@unal.edu.co}

\begin{document}
  \maketitle


\newcommand{\aad}{\langle a^\dagger\rangle}
\newcommand{\aaa}{\langle a\rangle}
\newcommand{\sss}{\langle \sigma\rangle}
\newcommand{\ssd}{\langle \sigma^\dagger\rangle}
\newcommand{\ada}{\langle a^\dagger a\rangle}
\newcommand{\ads}{\langle a^\dagger \sigma\rangle}
\newcommand{\sda}{\langle\sigma^\dagger a\rangle}
\newcommand{\sds}{\langle\sigma^\dagger\sigma\rangle}

\begin{resumen}
\noindent En este trabajo, se obtiene una expresión numérica para calcular el espectro de emisión de un sistema microcavidad-punto cuántico usando una teoría de campo medio en el formalismo de la matriz densidad. El sistema modelado es un micropilar semiconductor que contiene un único punto cuántico en el interior de la microcavidad, este sistema presenta pérdida de fotones a través de los espejos de la cavidad y bombeo de excitones al punto cuántico. Obtenemos una ecuación maestra de campo medio no lineal que nos permite calcular el espectro de emisión.
\end{resumen}


\begin{ingles}
\noindent In this work, we deduce a numerical expression to calculate the emission spectrum for a quantum dot-microcavity system, using a mean field theory in the density matrix formalism. The open system modeled, is a typical semiconductor micropillar that contains a single quantum dot inside a microcavity, this system presents leakage of photons through cavity mirrors and pumping of excitons to the quantum dot. We obtain a mean field nonlinear master equation that allow us calculate the emission spectrum.
\end{ingles}
\ikeywords{Microcavity, quantum dot, mean field theory}

\subsection*{Introducción}

\noindent Desde que Purcell descubrió el efecto que una cavidad tiene en el tiempo de relajación de un emisor óptico \citep{purcell}, lo emisores dentro de cavidades han sido de gran interés investigativo debido a sus potenciales aplicaciones tecnológicas, especialmente desde su implementación en sistemas de estado sólido como lo son, en particular, puntos cuánticos al interior de micropilares.\\

\noindent Es bien conocido que en estos sistemas se presentan dos regímenes: \textit{acople débil} y \textit{acople fuerte} (WC y SC, respectivamente, por sus siglas en inglés). En el primero, la interacción entre la radiación (modos ópticos de la cavidad) y la materia (puntos cuánticos) es tan débil que puede ser tratada perturbativamente; en el último se presenta una alta probabilidad de emisión y reabsorción de fotones por parte del emisor, generando estados altamente acoplados de luz y materia conocidos como \textit{estados vestidos}.\citep{laussy1}\\

\noindent Un modelo usualmente utilizado para la descripción de la física fundamental de este sistema es el de Jaynes Cummings ($\hbar=1$)\citep{gerry}

\begin{equation}
H=\omega_fa^\dagger a+\omega_a\sigma^\dagger\sigma+g(a^\dagger \sigma+\sigma^\dagger a)
\end{equation}

\noindent$a$ y $a^\dagger$ son los operadores de aniquilación y creación de fotones de la cavidad, respectivamente, con energía $\omega_f$, estos siguen la regla de conmutación $[a,a^\dagger]=1$; $\sigma$ y $\sigma^\dagger$ son los operadores escalera para excitón, además satisfacen la regla de anticonmutación $\{\sigma,\sigma^\dagger\}=1$ debido a la estadística fermiónica que le imponemos al punto cuántico de energía $\omega_a$; $g$ indica la fuerza de acople lineal entre radiación y materia.\\

\noindent El sistema que vamos a tratar incluye, además, una pérdida de fotones a través de los espejos de la cavidad, una pérdida de excitones debida a emisión espontánea y un bombeo incoherente de excitones al punto cuántico. Nos interesa analizar la validez de la aproximación de campo medio en la cual el valor esperado del producto de operadores se expresa como el producto de los valores esperados de cada operador\citep{vinck}. Luego de analizar la validez de la aproximación se propone un cálculo del espectro de emisión.\\

\noindent Fijamos $g=1\mathrm{meV}$ y lo tomamos como nuestra escala de energía\citep{tejedorpereaporras} a partir de la cual se determinan las demás cantidades.

\section*{Marco Teórico}

\noindent Estudios previos han mostrado la forma de tratar este sistema añadiendo términos disipativos de Lindblad a la ecuación maestra de Liouville-von Neumann para la evolución temporal de la matriz de densidad\citep{tejedorpereaporras}\citep{laussy2}

\begin{align}
\partial_t\rho=i[\rho,H]+\dfrac{\kappa}{2}(2a\rho a^\dagger-a^\dagger a\rho&-\rho a^\dagger a)+\dfrac{\gamma}{2}(2\sigma\rho\sigma^\dagger-\sigma^\dagger\sigma\rho-\rho\sigma^\dagger\sigma)\nonumber\\ &+\dfrac{P}{2}(2\sigma^\dagger\rho\sigma-\sigma\sigma^\dagger\rho-\rho\sigma\sigma^\dagger).
\label{ecuacion maestra}
\end{align}

\noindent El sistema pierde fotones con una tasa $\kappa$ a través de los espejos de la microcavidad, $\gamma$ es la tasa de decaimiento del excitón por emisión espontánea y $P$ es la tasa de bombeo continuo e incoherente de excitón al punto cuántico. Calculando los elementos de matriz en la base $(|Xn\rangle;|Gn\rangle)$ se obtiene $\rho_{in,jm}=\langle in|\rho|jm\rangle$, lo cual lleva al siguiente conjunto infinito de ecuaciones diferenciales\citep{tejedorpereaporras}:

\begin{align}
\partial_t\rho_{Gn,Gn}=&ig\sqrt{n}(\rho_{Gn,Xn-1}-\rho_{Xn-1,Gn})+\gamma\rho_{Xn,Xn}\nonumber\\
&-\kappa[n\rho_{Gn,Gn}-(n+1)\rho_{Gn+1,Gn+1}]-P\rho_{Gn,Gn}
\end{align}

\begin{align}
\partial_t\rho_{Xn,Xn}=&ig\sqrt{n+1}(\rho_{Xn,Gn+1}-\rho_{Gn+1,Xn})-\gamma\rho_{Xn,Xn}\nonumber\\
&-\kappa[n\rho_{Xn,Xn}-(n+1)\rho_{Xn+1,Xn+1}]+P\rho_{Gn,Gn}
\end{align}

\begin{align}
\partial_t\rho_{Gn,Xn-1}=&i[g\sqrt{n}(\rho_{Gn,Gn}-\rho_{Xn-1,Xn-1})-\Delta\rho_{Gn,Xn-1}]\nonumber\\& 
-[(\gamma+\kappa(2n-1)+P)/2] \rho_{Gn,Xn-1}\nonumber\\
&-\kappa\sqrt{n(n+1)}\rho_{Gn+1,Xn}
\end{align}

\begin{align}
\partial_t\rho_{Xn-1,Gn}=&-i[g\sqrt{n}(\rho_{Gn,Gn}-\rho_{Xn-1,Xn-1})-\Delta\rho_{Xn-1,Gn}]\nonumber\\& 
-[(\gamma+\kappa(2n-1)+P)/2] \rho_{Xn-1,Gn}\nonumber\\
&-\kappa\sqrt{n(n+1)}\rho_{Xn,Gn+1}
\end{align}

\noindent donde se ha definido el detuning $\Delta=\omega_f-\omega_a$. Una vez calculados los elementos de matriz, el número medio de fotones se puede obtener de:

\begin{equation}
N=\sum_mm(\rho_{Gm,Gm}+\rho_{Xm,Xm})
\end{equation}

\noindent Este tratamiento es lo que consideramos como "modelo exacto", ya que su planteamiento es analítico a pesar de que sea necesario truncar el conjunto de ecuaciones diferenciales; respecto a él compararemos nuestros resultados.\\

\noindent Teniendo en cuenta que el valor esperado de un operador se obtiene como $\langle O\rangle=Tr(\rho O)$ y la derivada del mismo en el cuadro de Schrödinger es $\dot{\langle O\rangle}=Tr(O\partial_t\rho)$, se calcula la dinámica de los valores esperados haciendo uso de \eqref{ecuacion maestra}. Al hacer esto, se obtienen valores esperados para los conjuntos de operadores $\langle\sigma^\dagger\sigma a^\dagger\rangle$ y $\langle a^\dagger a\sigma^\dagger\sigma\rangle$, los cuales en la aproximación de campo medio se expresan como:

\begin{equation}
\langle \sigma^\dagger\sigma a^\dagger\rangle\approx\sds\aad
\end{equation}

\begin{equation}
\langle a^\dagger a\sigma^\dagger\sigma\rangle\approx\aad\aaa\sds
\end{equation}

\noindent obteniendo las siguientes ecuaciones:

\begin{equation}
\dot{\aad}=i\omega_f\aad+ig\ssd-\dfrac{\kappa}{2}\aad
\label{ad}
\end{equation}

\begin{equation}
\dot{\ssd}=i\omega_a\ssd+ig\aad-2ig\aad\sds-\dfrac{\Gamma}{2}\ssd
\label{sd}
\end{equation}

\begin{equation}
\dot{\sds}=ig(\aad\sss-\ssd\aaa)-\Gamma\sds+P
\label{sds}
\end{equation}

\noindent y los hermíticos conjugados de \eqref{ad} y \eqref{sd}. Se ha definido $\Gamma=\gamma+P$ una tasa de disipación efectiva para el excitón. La razón para realizar la aproximación así y no, por ejemplo, como $\langle \sigma^\dagger\sigma a^\dagger\rangle\approx\ssd\sss\aad$, es porque de esta forma no se pierde información del bombeo $P$ el cual es importante para mantener las excitaciones en la cavidad, permitiendo la llegada a un estado estacionario, veamos. Si se tomara $\sds\approx\ssd\sss$, su derivada sería $\dot{\sds}\approx\dot{\ssd}\sss+\ssd\dot{\sss}$, con lo que se obtendría:

\begin{equation}
\dot{\sds}\approx ig(\aad\sss-\ssd\aaa)-\Gamma\ssd\sss
\end{equation}

\noindent Comparando con \eqref{sds} vemos que al separar $\sds$ se pierde parte del efecto que tiene el bombeo sobre el sistema.\\

\noindent Con la dinámica de los valores esperados resuelta, se define la función de correlación en la aproximación de campo medio, en analogía a lo realizado en \citep{laussy1}, como:

\begin{equation}
G^{mf}(t,\tau)=\langle a^\dagger (t)\rangle\langle a(t+\tau)\rangle
\label{correlacion}
\end{equation}

\noindent Con la función de correlación se puede calcular el espectro, en particular, se puede calcular para estados estacionarios que son en los que nos vamos a centrar:

\begin{equation}
S(\omega)=\dfrac{1}{M}\lim_{t\rightarrow\infty}\mathcal{R}\int_0^\infty G^{mf}(t,\tau)e^{i\omega\tau}d\tau
\end{equation}

\noindent siendo $M$ una constante de normalización.

\section*{Validez de la aproximación}

\noindent Al realizar la aproximación se pierde información de las reglas de conmutación debido a que al proponer $\ada\approx\aad\aaa$, los valores esperados del lado derecho conmutan, equivalente a decir que $[a,a^\dagger]=0$. Es de esperarse, entonces, que la aproximación dependa de las poblaciones medias de fotones en la cavidad, lo que la hace sensible a los parámetros $\kappa$ y $P$.\\

\begin{figure}
\begin{center}
\includegraphics[width=10cm]{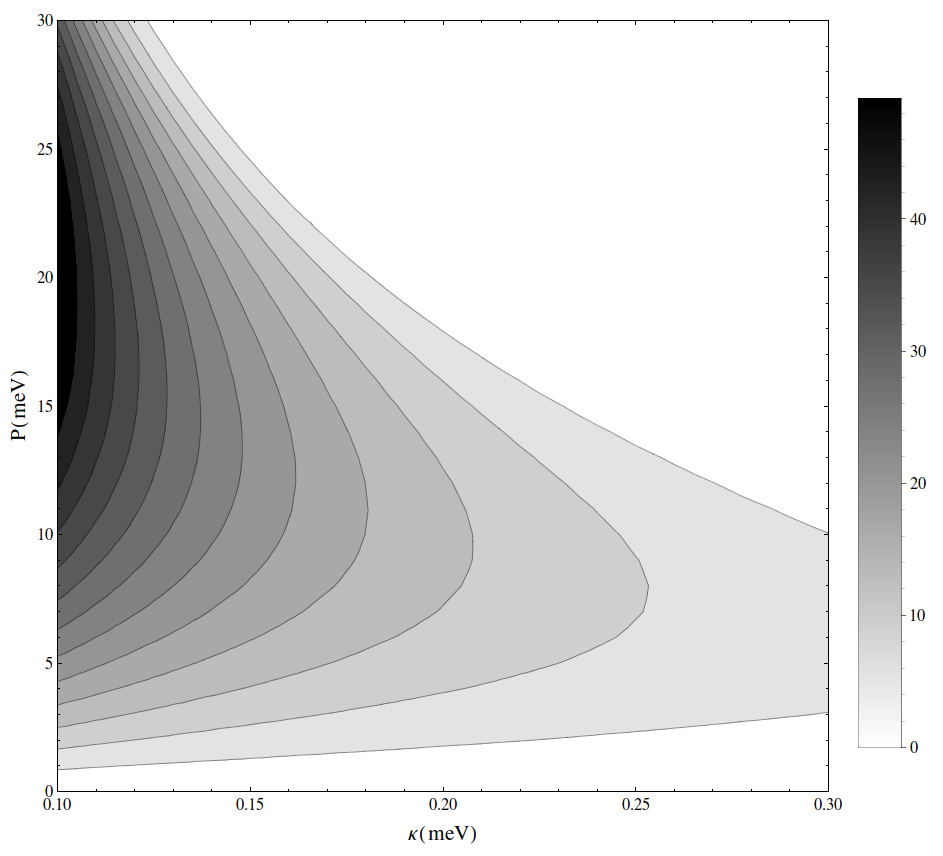}
\end{center}
\caption{\itshape Número medio de fotones en aproximación de campo medio $\bar{N}$ en escala de grises, como una función de $\kappa$ y $P$ en condiciones de resonancia $\Delta=0$ con $\gamma=0$.$1$ y $g=1$.}  
\label{numero fotones}
\end{figure}

\noindent Definimos el número medio de fotones en la aproximación de campo medio como:

\begin{equation}
\bar{N}=\aad\aaa
\end{equation}

\begin{figure}
\begin{center}
\includegraphics[width=10cm]{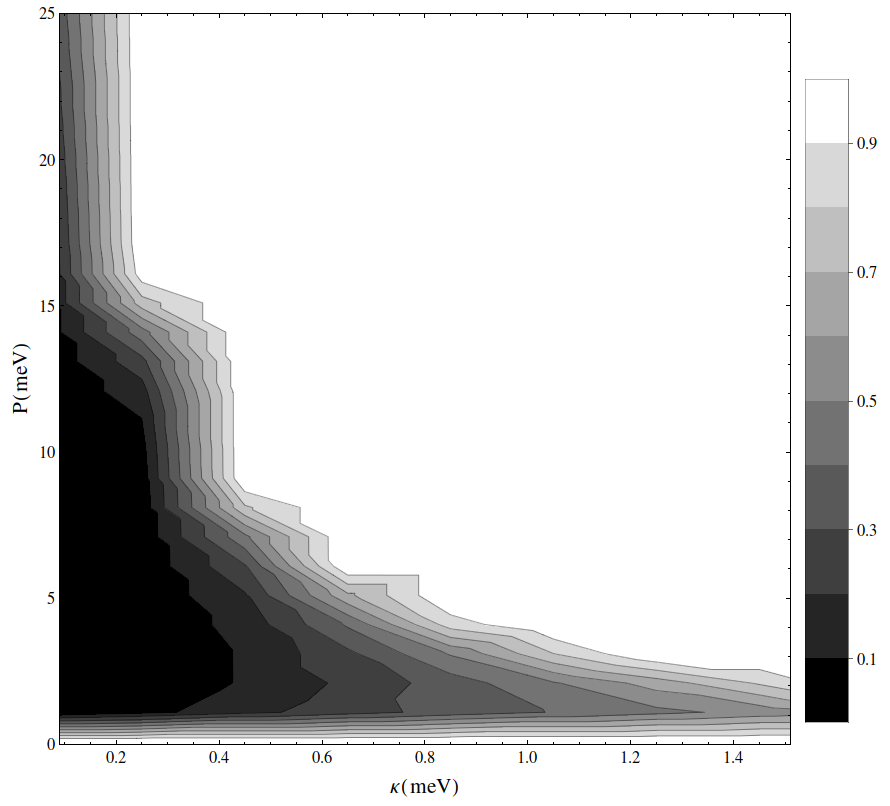}
\end{center}
\caption{\itshape Error relativo $\delta$, como una función de $\kappa$ y $P$ en condiciones de resonancia $\Delta=0$ con $\gamma=0$.$1$ y $g=1$, se observa un bajo error para pequeños valores de disipación.}  
\label{delta}
\end{figure}

\begin{figure}
\begin{center}
\includegraphics[width=12cm]{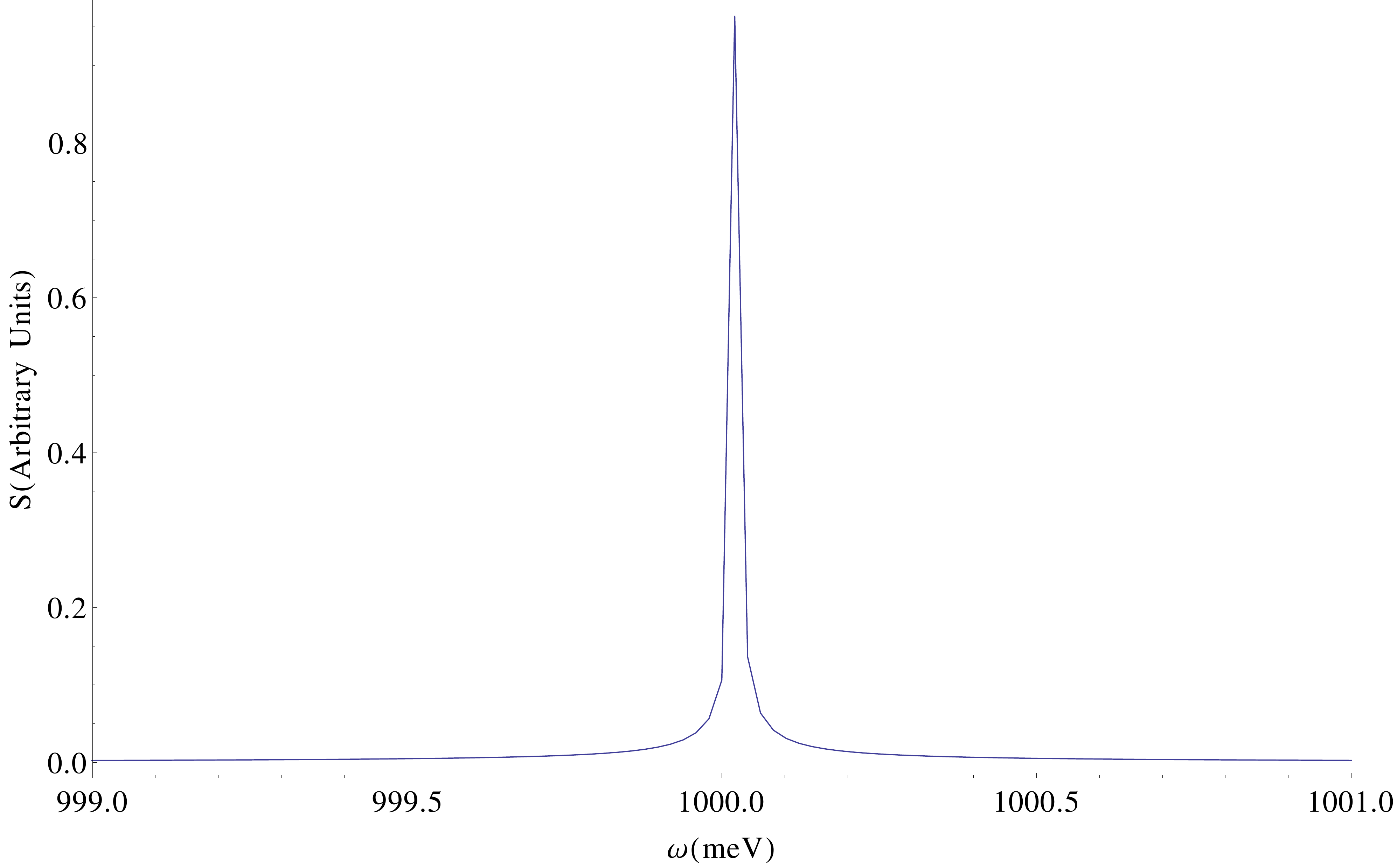}
\end{center}
\caption{\itshape Espectro de emisión del sistema en resonancia para los parámetros $\kappa=0$.$1$, $P=10$, $\gamma=0$.$1$ y $g=1$.}  
\label{espectro1}
\end{figure}

\begin{figure}

\begin{center}
\includegraphics[width=12cm]{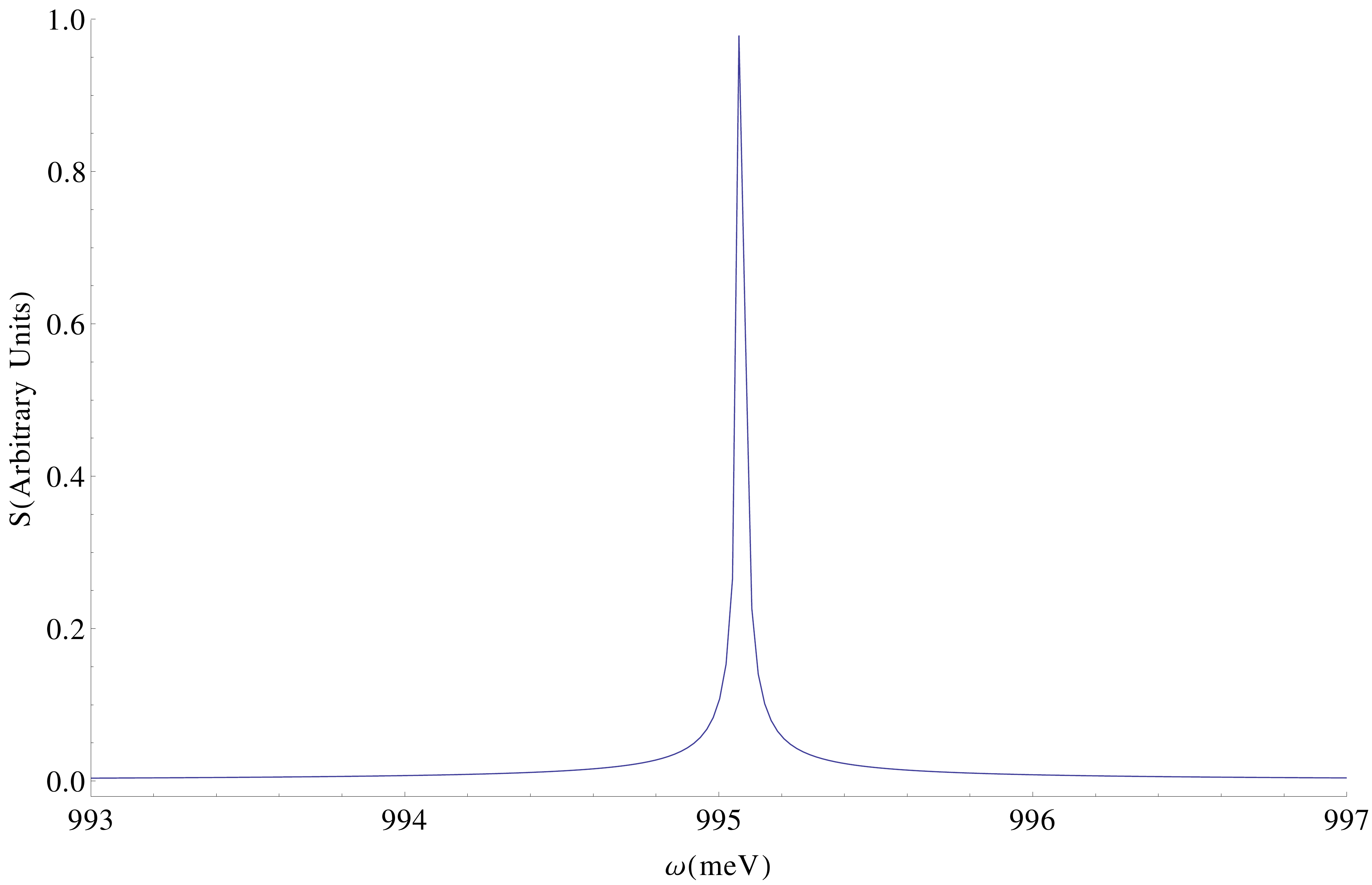}
\end{center}
\caption{\itshape Espectro de emisión del sistema con detuning $\Delta=-5$, $\kappa=0$.$1$, $P=10$, $\gamma=0$.$1$ y $g=1$.}  
\label{espectro2}
\end{figure}

\noindent La figura \ref{numero fotones} muestra cómo varía este último en función de $\kappa$ y $P$. Se obtiene un alto número de fotones para $\kappa\approx0$.$1$ y $P\approx20$, algo que es de esperarse ya que éste disminuye con su tasa de disipación y aumenta con el bombeo de excitaciones en la cavidad. Sin embargo, el bombeo del sistema se realiza en el excitón, de forma que valores de $P$ muy altos producen una saturación en la población de excitones, lo que impide el crecimiento de la población de fotones como lo muestra la figura, comportamiento que ya ha sido estudiado antes \citep{tejedorpereaporras}\citep{benson}\citep{mu}.\\

\noindent Con el objetivo de realizar un análisis cuantitativo de la validez de la aproximación de campo medio y de su sensibilidad a los parámetros disipativos, se define el error relativo $\delta=\left|(N-\bar{N})/N\right|$. La figura \ref{delta} evidencia el comportamiento de esta cantidad en función de $\kappa$ y $P$. Se encuentra una región bien definida en la que el error relativo es menor del 10\%. Vemos, además, que para grandes valores de $\kappa$, $\delta$ crece, esto es debido al bajo número medio de fotones que se presenta. Como estamos olvidando las reglas de conmutación, es de esperarse que la aproximación tenga mayor validez a medida que aumenta la población de fotones en la cavidad.

\section*{Espectro de Emisión}

\noindent Ya conociendo las regiones en que la aproximación es válida, nos remontamos al cálculo del espectro de emisión, que consiste en la transformada de Fourier de la función de correlación \eqref{correlacion}. El espectro obtenido, figura \ref{espectro1}, presenta su máximo pico en frecuencia $\omega\approx1000\mathrm{meV}$, algo que se esperaba debido a los valores tomados para las energías propias de la cavidad $\omega_{f,a}$. La figura \ref{espectro2} muestra el efecto que tiene el detuning. Los espectros fueron calculados para estado estacionario.

\subsection*{Conclusiones}

Se analizó cualitativa y cuantitativamente la aproximación de campo medio aquí descrita, para el cálculo del número medio de fotones y el espectro de emisión de un sistema microcavidad-punto cuántico. Se encontró que la aproximación es válida para  valores de $\kappa$ entre 0meV y 0.4meV, y para $P$ entre 1meV y 14meV, con un error relativo inferior al 10\%. Se calculó el espectro y éste reproduce el resultado conocido de reducción del ancho de línea para el caso de alto bombeo de excitones en la cavidad.

\subsection*{Agradecimientos}

Este trabajo ha sido financiado por Colciencias dentro del proyecto con código 110156933525, contrato número 026-2013 y código HERMES 17432. Por otra parte, reconocemos el apoyo técnico y computacional del Grupo de Óptica e Información Cuántica de la Universidad Nacional de Colombia, Sede Bogotá.

\bibliography{bibfile}

\end{document}